\documentclass[twocolumn,aps,prd,amsmath,superscriptaddress]{revtex4-1}
\usepackage{color}
\usepackage{fancyhdr}
\usepackage{graphicx}
\usepackage[ansinew]{inputenc}
\usepackage{amssymb}
\usepackage{amsmath}
\usepackage{cancel}
\usepackage{float}
\usepackage{graphicx}
\usepackage{float}
\usepackage{todonotes}
\usepackage{subfigure}
\usepackage{pgfplots}
\usepackage{silence}
\usepackage{lipsum}
\pgfplotsset{compat=1.9}
\usepackage{xparse}

\def\rklr#1{\left(#1\right)}

\begin{document}
\title{Implications of the fermion vacuum term in the extended SU(3) Quark Meson model on compact stars properties}
\author{Andreas Zacchi}
\email{zacchi@astro.uni-frankfurt.de}
\affiliation{Institut f\"ur Theoretische Physik, Goethe Universit\"at Frankfurt, 
Max von Laue Strasse 1, D-60438 Frankfurt, Germany}

\author{J\"urgen Schaffner-Bielich}
\email{schaffner@astro.uni-frankfurt.de}
\affiliation{Institut f\"ur Theoretische Physik, Goethe Universit\"at Frankfurt, 
Max von Laue Strasse 1, D-60438 Frankfurt, Germany}
\date{\today}
   \begin{abstract}

We study the impact of the fermion vacuum term in the SU(3) quark meson model on the equation of state and determine the vacuum parameters for various sigma meson masses. We examine its influence on the equation of state and on the resulting mass radius relations for compact stars. The tidal deformability $\Lambda$ of the stars is studied and compared to the results of the mean field approximation. 
Parameter sets which fulfill the tidal deformability bounds of GW170817 together with the observed two solar mass limit turn out to be restricted to a quite small parameter range in the mean field approximation. The extended version of the model does not yield solutions fulfilling both constraints.   
Furthermore, no first order chiral phase transition is found in the extended version of the model, not allowing for the twin star solutions found in the mean field approximation.

   \end{abstract}
        \maketitle
     \section{Introduction}
The theory of the strong interaction, quantum chromodynamics (QCD), describes the interaction between quarks and gluons. 
The QCD Lagrangian possesses an exact color- and flavor symmetry 
for $N_f$ massless quark flavours
\cite{Gasi69,Koch:1997ei,Karsch:2000kv,Zschiesche_review,Parganlija:2012gv,Parganlija:2012fy}  
and chiral symmetry controls the hadronic interactions 
in the low energy regime \cite{Gavin94a,Chandra99}. 
At high temperatures or densities chiral symmetry is 
expected to be restored \cite{Koch:1997ei,Kirzhnits:1972ut,Pisarski:1983ms}. 
The possible appearence of quark matter at high densities has interesting consequences for the properties of compact stars \cite{zdunikhaenselschaeffer:1983pti,Sagert:2008uq,SchaffnerBielich:2010am,Weissenborn:2011ut,Blaschke:2015uva,Zacchi:2015lwa,Zacchi:2015oma,Zacchi:2016tjw}.\\
Since QCD can not be solved on the lattice at nonzero density, effective models are needed to study the features and interactions of high density matter \cite{Toimela:1984xy,Mocsy:2004ab,Braaten96,Fraga:2004gz,Parganlija:2012gv,Parganlija:2016yxq}. 
The chiral SU(3) quark meson model is a well established and studied framework 
\cite{Kogut83,Koch95,Schaefer:2004en,Parganlija:2010fz,Parganlija:2012gv,Gupta:2011ez,Herbst:2013ufa,Stiele:2013pma,Lenaghan:1999si,Lenaghan:2000ey,Grahl:2011yk,Seel:2011ju,Parganlija:2016yxq}. Its advantage in comparison to other chiral models like the 
Nambu-Jona-Lasinio model 
\cite{Hara66,Bernard:1995hm,Schertler:1999xn,Buballa:2003qv} 
lies in its renormalizability \cite{Lenaghan:1999si,Lenaghan:2000ey,Zacchi:2017ahv} by taking into account vacuum fluctuations.\\

In this article we study the impact of the fermion vacuum term on the equation of state (EoS) in the SU(3) quark meson model in an extended mean field approximation (eMFA), see also \cite{Mocsy:2004ab,Fraga:2009pi,Skokov:2010sf,Gupta:2011ez,Chatterjee:2011jd,Tiwari:2013pg}. For that purpose we determine the vacuum parameters for different sigma meson masses $m_{\sigma}$. The dependence of the vacuum parameters on the renormalization scale parameter $\Lambda_{r}$ cancels with the dependence of the additional fermion vacuum term $\Omega_{\bar{q}q}^{vac}(\Lambda_{r})$ in the grand potential, so that 
the whole grand potential is independent on the renormalization scale parameter $\Lambda_{r}$  \cite{Lenaghan:1999si,Gupta:2011ez,Chatterjee:2011jd,Tiwari:2013pg}.
The resulting equations of state (EoSs) are investigated and subsequently used to solve the Tolman-Oppenheimer-Volkoff (TOV) equations \cite{Tolman39} for various parameters of the model, that is the sigma meson mass $m_{\sigma}$, the repulsive vector coupling constant $g_{\omega}$ and the vacuum pressure constant $B^{1/4}$. 
Constraints on the EoS for compact stars are imposed by the observation of the $2M_{\odot}$ neutron stars \cite{Demorest:2010bx,Antoniadis:2013pzd,Fonseca:2016tux,Cromartie:2019kug} and by the gravitational wave measurement GW170817 \cite{TheLIGOScientific:2017qsa,Abbott:2018wiz} of a binary neutron star merger. In this context, the tidal deformability parameter $\Lambda$ depends on the compactness $C$ of the compact star and on the Love number $k_2$ \cite{Love:1908tua,Hinderer:2007mb,Postnikov:2010yn} via 
\begin{equation}
 \Lambda=\frac{2 k_2}{3 C^5}.
\end{equation}
The GW170817 measurement on the tidal deformability deduces $\Lambda=300^{+420}_{-230}$ for a 1.4$M_{\odot}$ star \cite{Abbott:2018wiz}. Inferred from that measurement, the radius of a 1.4$M_{\odot}$ star cannot be larger than $R \geq 13.5$~km 
\cite{Abbott:2018exr,Most:2018hfd,Annala:2017llu,De:2018uhw,Kumar:2017wqp,Fattoyev:2017jql,Malik:2018zcf,Hornick:2018kfi}.\\
The mass radius relations of the eMFA are compared to those of the standard mean field approximation (MFA). We find that the mass radius relations for a given parameter set in the eMFA are in general less compact compared to the MFA case. Less compact star configurations imply rather large values of the tidal deformability parameter $\Lambda$.
Fulfilling all considered constraints on mass $M \geq 2M_{\odot}$, radius $R \leq 13.5$~km at 1.4$M_\odot$ and the tidal deformability parameter $\Lambda \leq 720$ for a 1.4$M_\odot$ star is possible in a narrow parameter space for the MFA. The parameters in the eMFA do not allow for solutions which satisfy the above mentioned constraints.\\ 
Further analysis of the parameter range of the SU(3) chiral quark meson model in the eMFA exhibits that the inclusion of the fermion vacuum term yields crossover transitions from a chirally broken phase to a restored phase exclusively. 
This feature consequently smoothens the EoSs compared to the MFA case 
not allowing for twin star solutions, that is two stable branches in the mass radius relation, as found for a certain parameter range in the MFA, see e.g. \cite{Zacchi:2016tjw}. 

\section{The SU(3) Quark Meson model}
A chirally invariant model with three flavours $N_f=3$ and with quarks as the active degrees of freedom  is the SU(3) chiral quark meson model.
Based on the theory of the strong interaction, an effective model must doubtless implement features of QCD, such as flavour symmetry and spontaneous- and explicit breaking of chiral symmetry. 
The $N_f=3$ Lagrangian \cite{Lenaghan:2000ey,Schaefer:2008hk,Parganlija:2012fy,Zacchi:2015lwa} 
respecting these symmetries and including vector mesonic interactions reads
\begin{eqnarray}\nonumber
\mathcal{L}&=&\sum_{\alpha}\bar{\Psi}_n\left(i\cancel{\partial}-g_{\alpha}{m_n}\right)\Psi_n+\bar{\Psi}_s\left(i\cancel{\partial}-g_{\alpha}{m_s}\right)\Psi_s\\ \nonumber
&+&\rm{tr}(\partial_{\mu}\Phi)(\partial^{\mu}\Phi)^{\dagger}-\lambda_1[tr(\Phi^{\dagger}\Phi)]^2-\lambda_2 tr(\Phi^{\dagger}\Phi)^2\\ \nonumber
&+&
tr[\hat{H}(\Phi+\Phi^{\dagger})]+c\left(\det(\Phi^{\dagger})+\det(\Phi)\right)\\
&-&m_0^2(tr(\Phi^{\dagger}\Phi))-m_v^2 tr(V^{\dagger}V) \label{lagrangiandensityofthesuthreecasebaby}
\end{eqnarray}
with a Yukawa like coupling $g_{\alpha}$, $\alpha$ being the fields $\sigma_n$, $\sigma_s$, $\omega$, $\rho$ and $\phi$ involved, and the effective mass $m_{n,s}$ to the spinors $\Psi_{n,s}$. The indices n$=$nonstrange and s$=$strange indicate the flavour content.  
All physical fields are arranged in the matrix $\Phi$ \cite{Parganlija:2012gv,Parganlija:2012fy}
\small{
\begin{eqnarray}\nonumber
\Phi=\frac{1}{\sqrt{2}}
\left(\begin{array}{ccc}
\frac{(\sigma_n+a_0^0)+i(\eta_n+\pi_0)}{\sqrt{2}}&a_0^{+}+i\pi^+&K_s^{+}+iK^+\\
a_0^{-}+i\pi^-&\frac{(\sigma_n-a_0^0)+i(\eta_n-\pi^0)}{\sqrt{2}}&K_s^0+iK^0\\
K_s^-+iK^-&\bar{K}_s^0+i\bar{K}^0&\sigma_s+i\eta_s\\
\end{array}\right) \\\label{mmmaaatttrrriks_alibaba}
\end{eqnarray}}
 \normalsize
where we consider the condensed $\sigma_{n,s}$ fields and the pions to determine the vacuum parameters
$\lambda_1$, $\lambda_2$, $m_0^2$, c, 
$h_n$ and $h_s$ of the model, which are fixed at tree level \cite{Schaefer:2008hk,Lenaghan:2000ey,Parganlija:2010fz,Beisitzer:2014kea}, but change upon renormalization \cite{Chatterjee:2011jd,Tiwari:2013pg,Zacchi:2017ahv}, see Sec.\ref{renormalization}.
In thermal equilibrium the grand potential $\Omega$ is calculated via the partition 
function $\mathcal{Z}$, which is defined as a path integral over the fermion fields. 
\begin{eqnarray}\label{oho}
\Omega=-\frac{\ln\mathcal{Z}}{\beta} \quad\hbox{with}\,\, \mathcal{Z}=\int \mathcal{D}\Psi\mathcal{D}\bar{\Psi}e^{\int_0^{\beta}d\tau \int d^3\vec{x} \mathcal{L}} 
\end{eqnarray}
Evaluated, the grand canonical potential reads 
\begin{eqnarray} \nonumber
\Omega_{\bar{q}q}&=&V  + \Omega_{\bar{q}q}^{vac} + \Omega_{\bar{q}q}^{th}\\ \label{dumblidorr}
      &=&V-\frac{3}{\pi^2\beta}\int^{\infty}_0 k^2dk \cdot \left(\mathcal{R} + \mathcal{N}\right)
\end{eqnarray}
where V is the tree level potential 
\begin{eqnarray} \nonumber
V&=&\frac{\lambda_1}{4}\left((\sigma_n^2+\sigma_s^2)^2\right)+\frac{\lambda_2}{8}\left(\sigma_n^4+2\sigma_s^4\right)+\frac{m_0^2}{2}(\sigma_n^2 + \sigma_s^2)\\ \nonumber
&-&h_n\sigma_n-h_s\sigma_s-\frac{c\sigma_n^2\sigma_s}{2\sqrt{2}}\\ \label{tree_level_baby}
&-&\frac{1}{2}\left(m_{\omega}^2\omega^2+m_{\rho}^2\rho^2+m_{\phi}^2\phi^2\right)+B^{1/4}
\end{eqnarray}
$B^{1/4}$ is a phenomenological vacuum pressure term \cite{Bodmer:1971we,Chodos74,Farhi:1984qu,Witten:1984rs}, 
which will stiffen or soften the EoS $p(\epsilon)$, pressure $p$ vs. energy density $\epsilon$.
The fermion vacuum term is 
\begin{equation}\label{vacuum_level_baby}
 \Omega_{\bar{q}q}^{vac}=\mathcal{R}=\frac{E_{n,s}}{T},
\end{equation}
and the part from the mean field approximation is 
\begin{eqnarray}\nonumber
\Omega_{\bar{q}q}^{th}&=&\mathcal{N}=\ln\left(1+e^{-\beta(E_n+\tilde\mu_u)}\right)+\ln\left(1+e^{-\beta(E_n-\tilde\mu_u)}\right)\\ \nonumber
&+&\ln\left(1+e^{-\beta(E_n+\tilde\mu_d)}\right)+\ln\left(1+e^{-\beta(E_n-\tilde\mu_d)}\right)\\ \label{ehm}
&+&\ln\left(1+e^{-\beta(E_s+\tilde\mu_s)}\right)+
\ln\left(1+e^{-\beta(E_s-\tilde\mu_s)}\right)
\end{eqnarray}
The dependence of $\Omega_{\bar{q}q}^{th}$ on the chiral condensates $\sigma_{n,s}$ is 
implicit in the relativistic quasi-particle
dispersion relation for the constituent quarks 
\begin{equation}
E_{n,s}=\sqrt{k^2+\tilde{m}_{n,s}^2} 
\end{equation}
The quantity $\tilde{m}_{n,s}=g_{\alpha}m_{n,s}$ is the in medium mass generated by the scalar 
fields. $\mu_{n,s}$ in eq.~(\ref{ehm}) are the respective chemical potentials
\begin{eqnarray}
\tilde{\mu}_{up}&=&\tilde{\mu}_{u}=g_{\omega}\omega+g_{\rho}\rho \\
\tilde{\mu}_{down}&=&\tilde{\mu}_{d}=g_{\omega}\omega-g_{\rho}\rho\\ \label{eq:kichererbsen}
\tilde{\mu}_{strange}&=&\tilde{\mu}_{s}=g_{\phi}\phi 
\end{eqnarray}
%
\subsection{Renormalized vacuum parameters of the SU(3) Quark Meson model}\label{renormalization}
The implementation of the fermion vacuum term needs regularization schemes. 
In this work we employ the minimal substraction scheme and follow the procedure as found 
in \cite{Mocsy:2004ab,Fraga:2009pi,Skokov:2010sf,Gupta:2011ez,Chatterjee:2011jd,Tiwari:2013pg,Zacchi:2017ahv} to proper perform the regularization of the divergence. The diverging integral containing the fermion vacuum contribution in eq.~(\ref{dumblidorr}) is to lowest order just the one-loop
effective potential at zero temperature \cite{Skokov:2010sf} and is dimensionally regularized via the corresponding counter 
term 
\begin{equation}
 \delta \mathcal{L}=\frac{N_c N_f}{16\pi^2}\tilde{m}_{n,s}^4\left[\frac{1}{\epsilon}-\frac{1}{2}\left[-3+2\gamma-4 \ln \left(2\sqrt{\pi}\right)\right]\right]. 
\end{equation}
which gives
\begin{equation}
\Omega_{\bar{q}q}^{vac}=\frac{N_c N_f}{16\pi^2}\tilde{m}_{n,s}^4\left[\frac{1}{\epsilon}-\frac{1}{2}\left[-3+2\gamma+4 \ln \left(\frac{\tilde{m}_{n,s}}{2\sqrt{\pi}\Lambda_r}\right)\right]\right],
\end{equation}
where $N_c=3$ is the number of colors, $\lim\limits \epsilon  \rightarrow 0$ from dimensional reasoning, $\gamma$ is the Euler-Mascheroni constant and $\Lambda_r$ the renormalization scale parameter. 
The dimensionally regularized fermion vacuum contribution eventually reads 
\begin{equation} \label{yellasaidthe3}
\Omega_{\bar{q}q}^{dr}=-\frac{N_c N_f}{8\pi^2}\tilde{m}_f^4 \ln\rklr{\frac{\tilde{m}_{n,s}}{\Lambda_r}} 
\end{equation}
As in mean field approximation, the six model parameters 
$\lambda_1$, $\lambda_2$, $m_0^2$, c, $h_n$ and $h_s$ are 
fixed by six experimentally known values \cite{Lenaghan:2000ey,Schaefer:2008hk,Agashe:2014kda}. 
As an input the pion mass $m_{\pi}=136$~MeV, the kaon mass $m_{k}=496$~MeV, 
the pion decay constant $f_{\pi}=92.4$~MeV, the kaon decay constant 
$f_k=113$~MeV, the masses of the eta  meson $m_{\eta}=548$~MeV, the mass 
of the eta-prime meson $m_{\eta}'=958$~MeV and the mass of the sigma 
meson $m_{\sigma}$ need to be known.\\ 
The mass of the sigma meson $m_{\sigma}$ is experimentally not well 
determined. Usually, the sigma meson is identified with the experimentally 
measured resonance $\rm{f_0(500)}$, which is rather broad, 
$400 \leq m_{f_0} \leq 600$~MeV \cite{Ishida98,Agashe:2014kda}. 
In Ref.~\cite{Parganlija:2012fy} it was demonstrated that 
within an extended quark-meson model that includes vector and 
axial-vector interactions, the resonance $\rm{f_0(1370)}$ 
could be identified as this scalar state. 
The mass of the sigma meson is chosen from $400 \leq m_{\sigma} \leq  800$~MeV in the following.\\
Starting point for the determination of the vacuum parameters is the potential V,  eq.~(\ref{tree_level_baby}), including the pions 
and the vacuum contribution $\Omega_{\bar{q}q}$, eq.~(\ref{vacuum_level_baby}), which together form $\mathcal{V}$.
\begin{eqnarray} \nonumber
\mathcal{V}&=&\frac{\lambda_1}{4}\left[(\sigma_n^2+\sigma_s^2)^2+2\pi_0^2(\sigma_n^2+\sigma_s^2)+\pi_0^4 \right]\\ \nonumber
&+&\frac{\lambda_2}{8}\left[(\sigma_n^2+\pi_0^2)^2+2\sigma_s^4\right]+\frac{m_0^2}{2}(\sigma_n^2+\pi_0^2+\sigma_s^2)\\ \nonumber
&-& h_n\sigma_n-h_s\sigma_s - c\left(\frac{\sigma_n^2\sigma_s+\pi_0^2\sigma_s}{2\sqrt{2}}\right)+B^{1/4}\\ \nonumber
&+&\frac{1}{2}\left(m_{\omega}^2\omega^2+m_{\rho}^2\rho^2+m_{\phi}^2\phi^2\right)\\ \label{gantsestollespotenzziahhl}
&-&\frac{N_c N_f}{8\pi^2}\left(\sigma_n^2+\sigma_s^2 \right)^2 \ln\rklr{\frac{g\sqrt{\sigma_n^2+\sigma_s^2}}{\Lambda_r}}
\end{eqnarray}
The procedure to determine the vacuum parameters is similar as in the SU(2) case \cite{Zacchi:2017ahv}. Here, three of the six derivatives 
to determine the vacuum paramters read 
\begin{eqnarray}\nonumber
 h_n&=&m_0^2\sigma_n-\frac{c\sigma_s\sigma_n}{\sqrt{2}}+\lambda_1\left(\sigma_n^2+\sigma_s^2\right)\sigma_n+\frac{\lambda_2}{2}\sigma_n^3 +\rm{Vac.}\\ \nonumber
 h_s&=&m_0^2\sigma_s-\frac{c\sigma_n^2}{\sqrt{2\sqrt{2}}}+\lambda_1\left(\sigma_n^2+\sigma_s^2\right)\sigma_s+\lambda_2\sigma_s^3 +\rm{Vac.} \\ \nonumber
 m_{\pi}^2&=&m_0^2-\frac{c\sigma_s}{\sqrt{2}}+\lambda_1\left(\sigma_n^2+\sigma_s^2\right)+\frac{\lambda_2}{2}\sigma_n^2+\rm{Vac.}\\
 \label{drei}
\end{eqnarray}
with the vacuum contribution yet to be determined.
Since the pion does not form a condensate, it does not 
occur anymore in the equations~(\ref{drei}) above. 
Due to mixing of the fields, the second derivative $\partial^2 \mathcal{V} / \partial \sigma_n^2$ 
does not yield the mass of the sigma meson as in the two flavour case SU(2). 
At this point the kaon mass is 
needed as input parameter and one remains with three unknown quantities \cite{Gupta:2011ez}.
%
It is necessary to rewrite the nonstrange-strange basis 
in terms of the generators, i.e. the mathematical fields and afterwards idenitfying  
those with the physical fields. 
Following \cite{Lenaghan:2000ey,Schaefer:2008hk} the matrix 
$\Phi$, eq.~(\ref{mmmaaatttrrriks_alibaba}), can be written as 
$\Phi=T_a\Phi_a=T_a(\sigma_a + i\pi_a)$ with $T_a=\frac{\lambda_a}{2}$,  
$\lambda_a$ being the Gell Mann matrices where $a=0,1,2,..,8$ are the 
nine generators of the U(3) symmetry group. The generators 
obey the U(3) algebra with the standard symmetric $d_{abc}$ and 
antisymmetric structure constants $f_{abc}$. Rearranging the 
entries of eq.~(\ref{mmmaaatttrrriks_alibaba}) gives
\small{
\begin{eqnarray}\nonumber
\Phi=\frac{1}{2}
\left(\begin{array}{ccc}
\sqrt{\frac{2}{3}}\sigma_0 + \sigma_3+\frac{\sigma_8}{\sqrt{3}}&\sigma_1 - i\sigma_2&\sigma_4-i\sigma_5\\
\sigma_1+i\sigma_2&\sqrt{\frac{2}{3}}\sigma_0 - \sigma_3+\frac{\sigma_8}{\sqrt{3}}&\sigma_6-i\sigma_7\\
\sigma_4+i\sigma_5&\sigma_6+i\sigma_7&\sqrt{\frac{2}{3}}\sigma_0-\frac{2\sigma_8}{\sqrt{3}}\\
\end{array}\right)\\ \label{mmmaaatttrrriks_generators}
\end{eqnarray}}
\normalsize
and the transformation from the physical nonstrange-strange 
basis to the mathematical basis reads  
\begin{equation}\label{hinundherdasistschonschwer}
\left(\begin{array}{c}
\sigma_n\\
\sigma_s\\
\end{array}\right)
=\frac{1}{\sqrt{3}}
\left(\begin{array}{cc}
\sqrt{2}&1\\
1&-\sqrt{2}\\
\end{array}\right)
\left(\begin{array}{c}
\sigma_0\\
\sigma_8\\
\end{array}\right).
\end{equation}
Separating the entries for the scalar and the pseudoscalar
sector gives the potential in terms of the mathematical 
fields \cite{Lenaghan:2000ey,Schaefer:2008hk}.
The mass matrix $m_{ij}$  is determined only by the mesonic part 
and by the fermionic vacuum term of the potential $\mathcal{V}$, eq.~(\ref{gantsestollespotenzziahhl}),  
because the quark contribution vanishes at $T=\mu=0$. 
Because of isospin symmetry some entries of $m_{ij}^2$ are degenerate and 
furthermore $m_{08}^2=m_{80}^2$, so that  
\begin{equation}\label{mmmatriques}
 m_{ij}^2=\frac{\partial^2\mathcal{V}}{\partial\Phi_i \partial \Phi_j}=
\left(\begin{array}{ccc}
m_{00}^2&\dots&m_{08}^2\\
\vdots&\ddots    &\vdots\\
m_{80}^2&\dots&m_{88}^2\\
\end{array}\right)
\end{equation}
This matrix needs to be diagonalized for $m_{\sigma}^2$ 
and $m_{f_0}^2$ in the scalar sector, and for $m_{\eta}$ 
and $m_{\eta}'$ in the pseudoscalar sector introducing a mixing 
angle $\theta$. 
Eventually the mass of the kaon in the nonstrange-strange basis reads
\begin{equation}
 m_k^2=m_0^2-\frac{c\sigma_n}{2}+\lambda_1\left(\sigma_n^2+\sigma_s^2\right)+\frac{\lambda_2}{2}\left(\sigma_n^2-\sqrt{2}\sigma_n\sigma_s+2\sigma_s^2\right)
\end{equation}
which determines the axial anomaly term to be  
\begin{equation}\label{you_no_poo}
 c=\frac{-2\left(m_k^2-m_{\pi}^2\right)-\lambda_2\left( \sqrt{2}\sigma_n\sigma_s-2\sigma_s^2\right)}{\sigma_n-\sqrt{2}\sigma_s}.
\end{equation}
Using eq.~(\ref{mmmatriques}), 
the sum of $m_{\eta}$ and $m_{\eta}'$ reads
\begin{eqnarray}\nonumber 
 m_{\eta}+m_{\eta}'&=&2m_0^2+2\lambda_1\left(\sigma_n^2+\sigma_s^2\right)+\frac{\lambda_2}{2}\left(\sigma_n^2+2\sigma_s^2\right)+\frac{c \sigma_s}{\sqrt{2}}\\ \label{you_know_whos_the_poo}
 &=&2m_{\pi}^2-\frac{\lambda_2}{2}\left(\sigma_n^2-2\sigma_s^2\right)+\frac{3c\sigma_s}{\sqrt{2}}, 
\end{eqnarray}
and inserting eq.~(\ref{you_no_poo}) in eq.~(\ref{you_know_whos_the_poo}) 
to solve for $\lambda_2$ gives
\begin{equation}
 \lambda_2=\frac{m_{\eta}^2+m_{\eta}'^2-2m_{\pi}^2 + \frac{6\sigma_s(m_k^2-m_{\pi}^2)}{\sqrt{2}(\sigma_n-\sqrt{2}\sigma_s)}}  {\sigma_s^2-\frac{\sigma_n^2}{2} -   \frac{(3\sqrt{2}\sigma_s^2\sigma_n-6\sigma_s^3)}{\sqrt{2}(\sigma_s-\sqrt{2}\sigma_s)}  }
\end{equation}
The further procedure in the mean field approximation is to determine 
$\lambda_1(m_0^2)$ via $m_{\sigma}$ and $m_{\pi}$ \cite{Lenaghan:2000ey,Schaefer:2008hk,Zacchi:2015lwa}. The quantities obtained so far 
enter into the two condensate 
equations, eqs.~(\ref{drei}), for the explicity symmetry breaking terms $h_n$ and $h_s$. 
Working in the extended version of the model, 
the vacuum contributing part from eq.~(\ref{gantsestollespotenzziahhl})
has to be rewritten in terms of the mathematical fields. 
Furthermore 
\begin{eqnarray}\label{gold_1}
 \frac{\partial^2 \Omega_{\bar{q}q}^{dr}}{\partial^2
 \sigma_0^2}&=&\kappa\left[\mathcal{A}\mathcal{W}
 +\frac{\mathcal{X}\left(\frac{8}{3}\sigma_0-\frac{2\sqrt{2}}{3}\sigma_8\right)}{\mathcal{Z}}\right]\\ \label{gold_2}
 \frac{\partial^2 \Omega_{\bar{q}q}^{dr}}{\partial^2
 \sigma_0\sigma_8}&=&\kappa\left[\mathcal{B}\mathcal{W}
 +\frac{\mathcal{X}\left(\frac{10}{3}\sigma_8-\frac{2\sqrt{2}}{3}\sigma_0\right)}{\mathcal{Z}}\right]\\ \label{gold_3}
 \frac{\partial^2 \Omega_{\bar{q}q}^{dr}}{\partial^2
 \sigma_8^2}&=&\kappa\left[\mathcal{C}\mathcal{W}
 +\frac{\mathcal{Y}\left(\frac{10}{3}\sigma_8-\frac{2\sqrt{2}}{3}\sigma_0\right)}{\mathcal{Z}}\right]
\end{eqnarray}
where 
\begin{eqnarray}
 \kappa&=&-\frac{N_cN_f}{72\pi^2}g^4\\
\mathcal{A}&=&96\sigma_0^2-48\sqrt{2}\sigma_0\sigma_8+48\sigma_8^2\\
\mathcal{B}&=&96\sigma_0\sigma_8-30\sqrt{2}\sigma_8^2-24\sqrt{2}\sigma_0^2\\
\mathcal{C}&=&48\sigma_0^2-60\sqrt{2}\sigma_8\sigma_0+150\sigma_8^2\\
\mathcal{W}&=&2\ln\left(\frac{g\sqrt{\frac{4}{3}\sigma_0^2+\frac{2\sqrt{2}}{3}\sigma_0\sigma_8+\frac{5}{3}\sigma_8^2}}{\Lambda}\right)+\frac{1}{2}\\ \mathcal{X}&=&32\sigma_0^3-10\sqrt{2}\sigma_8^3-24\sqrt{2}\sigma_0^2\sigma_8+48\sigma_8^2\sigma_0\\
\mathcal{Y}&=&50\sigma_8^3-8\sqrt{2}\sigma_0^3+48\sigma_0^2\sigma_8-30\sqrt{2}\sigma_0\sigma_8^2\\
\mathcal{Z}&=&\frac{4}{3}\sigma_0^2-\frac{2\sqrt{2}}{3}\sigma_0\sigma_8+\frac{5}{3}\sigma_8^2.
\end{eqnarray}
Rewriting everything according to eq.~(\ref{hinundherdasistschonschwer}) 
in terms of the physical fields, these mass corrections enter in the 
numerical routine to search for the vacuum parameters $\lambda_1$ and $m_0^2$. 
These two parameters alone compensate the vacuum contribution in the two 
condensate equations $h_n$ and $h_s$ in eqs.~(\ref{drei}).\\
It is interesting to note that the grand canonical potential remains 
unaffected by the choice of the renormalization scale parameter $\Lambda_r$. 
This is easily seen in the SU(2) case \cite{Zacchi:2017ahv} and has also  
been shown for the SU(3) case for $m_{\sigma}=400$~MeV \cite{Chatterjee:2011jd,Gupta:2011ez}. To compare our results with \cite{Chatterjee:2011jd}, the renormalization scale parameter is set to $\Lambda_r=200$~MeV. 
\begin{table*}
\begin{center}
\begin{tabular}{|c|c|c|c|c|c|c|c|}
\hline\hline 
 $m_{\sigma}^{vac}$ & $\Lambda$ & $\lambda_1$  & $\lambda_2$ & $c [MeV]$  & $m^2 [MeV^2]$ & $h_n [MeV^3]$ & $h_s [MeV^3]$\\
\hline
 400 & - & -5.901 & 46.488 & 4807.245 & $(494.272)^2$ &$(120.73)^3$&$(336.41)^3$\\
\cline{2-8}
 500 & - & -2.698 & 46.488 & 4807.245 & $(434.541)^2$ &$(120.73)^3$& $(336.41)^3$ \\
\cline{2-8}
600 & - & 1.398 & 46.488 & 4807.245 & $(342.496)^2$   &$(120.73)^3$&$(336.41)^3$\\ 
\cline{2-8}
 700 & - & 6.615 & 46.488 & 4807.245 & $(161.918)^2$  &$(120.73)^3$& $(336.41)^3$\\ 
\cline{2-8}
 800 & - & 13.488 & 46.488 & 4807.245 & $-(306.289)^2$   &$(120.73)^3$&$(336.41)^3$\\ 
\cline{2-8}
 900 & - & 23.649 & 46.488 & 4807.245 & $-(520.82)^2$ & $(120.73)^3$& $(336.41)^3$ \\ 
\hline \hline \hline
 $m_{\sigma}^{vac}$ & $\Lambda$ & $\lambda_1$  & $\lambda_2$ & $c [MeV]$  & $m^2 [MeV^2]$ & $h_n [MeV^3]$ & $h_s [MeV^3]$\\
\hline
400 & 200 & -8.173 & 138.516 & 4807.245 & $(283.901)^2$ & $(120.73)^3$  & $(336.41)^3$ \\
\cline{2-8}
  500 & 200 & -5.285 & 138.516 & 4807.245 & $(173.694)^2$ &$(120.73)^3$&  $(336.41)^3$\\
\cline{2-8}
 600 & 200 & -1.661 & 138.516 &    4807.245& $-(181.951)^2$ &$(120.73)^3$&$(336.41)^3$\\ 
\cline{2-8}
 700 & 200 & 2.819 & 138.516 &  4807.245 & $-(333.675)^2$&$(120.73)^3$& $(336.41)^3$\\ 
\cline{2-8}
 800 & 200 & 8.450 & 138.516 &    4807.245& $-(457.902)^2$&$(120.73)^3$&$(336.41)^3$\\ 
\cline{2-8}
 900 & 200 & 16.179 & 138.516 &  4807.245& $-(587.056)^2$ &$(120.73)^3$&  $(336.41)^3$\\ 
\hline\hline
\end{tabular}
\caption{\textit{The vacuum parameters $\lambda_1$, $\lambda_2$, c, $m^2$, $h_n$ and $h_s$ for different values of the sigma meson mass $m_{\sigma}^{vac}$ in mean field approximation (MFA-upper table) and with the inclusion of the fermion vacuum term for the renormalization scale parameter $\Lambda=200$~MeV (eMFA-lower table).}}
\label{wirsindallebluna}
\end{center}
\end{table*}
%
\subsection{Charge neutrality and the Gap equations}\label{elefantige_leptonen}
 Since the lepton contribution decouples from the quark grand canonical potential it can be treated separately. The electron contribution reads 
\begin{equation}\label{gancanlegants}
\Omega_e=-\frac{2}{\beta}\int\frac{d^3k}{(2\pi)^3}\ln\left(1+e^{-\frac{E_{k,e}\pm\mu_e}{T}}\right)
\end{equation}
with $E_{k,e}=\sqrt{k^2+m_e^2}$, where $m_e$ is the electron mass and $\mu_e$ is the electron chemical potential.\\
A compact star is charge neutral, so that 
\begin{equation}\label{ladungnullinger}
\sum_{f=u,d,s,e}Q_f n_f=\frac{2}{3}n_u-\frac{1}{3}n_d-\frac{1}{3}n_s-n_e=0
\end{equation}
with $n_f$ as the particle density of each species f.\\
The total grand canonical potential is
\begin{eqnarray}\nonumber
 \Omega^{tot}&=&V-\frac{N_c N_f}{8\pi^2}\tilde{m}_f^4 \ln\rklr{\frac{\tilde{m}_f}{\Lambda}}-\frac{3}{\pi^2\beta}\int^{\infty}_0 k^2dk \cdot \mathcal{N}\\ \label{dasgantseverdammteding}
 &-&\frac{2}{\beta}\int\frac{d^3k}{(2\pi)^3}\ln\left(1+e^{-\frac{E_{k,e}\pm\mu_e}{T}}\right),
\end{eqnarray}
and the vacuum parameter sets for different sigma meson mass $m_{\sigma}$ are listed in Tab.~\ref{wirsindallebluna}. The equations of motion 
\begin{equation}\label{condensates}
\frac{\partial\Omega^{tot}}{\partial\sigma_n}=\frac{\partial\Omega^{tot}}{\partial\sigma_s}=\frac{\partial\Omega^{tot}}{\partial\omega}=\frac{\partial\Omega^{tot}}{\partial\rho}=\frac{\partial\Omega^{tot}}{\partial\phi}\overset{!}{=}0,\\
\end{equation}
also known as the gap-equations, finally determine the EoS, $p(\epsilon)$, with the pressure $p$ and the energy density $\epsilon$.
\section{Compact stars}\label{peppa_wutz}
The formalism discussed in the previous section can be used to obtain EoSs, $p(\epsilon)$, applicable for the calculation of compact stars. The resulting mass radius relations have to respect certain  constraints to be physically reasonable. The most important constraints are the $2M_{\odot}$ limit \cite{Demorest:2010bx,Antoniadis:2013pzd,Fonseca:2016tux,Cromartie:2019kug} and the tidal deformability measurement by the LIGO/Virgo collaboration \cite{TheLIGOScientific:2017qsa,Abbott:2018wiz}. 
\subsection{Tidal deformability}\label{gandalf}
The observation of the binary neutron star merger event GW170817 \cite{TheLIGOScientific:2017qsa} 
is used to constrain the EoSs of compact stars \cite{Radice:2017lry,Margalit:2017dij,Rezzolla:2017aly,Abbott:2018exr,Most:2018hfd,Annala:2017llu,De:2018uhw,Kumar:2017wqp,Fattoyev:2017jql,Malik:2018zcf}.
During the inspiral phase, one stars quadrupole deformation $Q_{ij}$ in response to the companions perturbing tidal field $\mathcal{E}_{ij}$ is measured by the tidal polarizability $\lambda$ 
\begin{equation}
Q_{ij}=-\lambda \mathcal{E}_{ij} .
\end{equation}
$\lambda$ depends on the EoS \cite{Hinderer:2007mb,Hinderer:2009ca,Postnikov:2010yn} and is related to the stars quadrupolar tidal Love number \cite{Love:1908tua} 
$k_2$ via 
\begin{equation}\label{liability}
k_2=\frac{3}{2}\lambda R^{-5} \, ,
\end{equation}
where $R$ is the radius of the star.\\
The Love number $k_2$ is calculated as follows  
\begin{eqnarray} \label{loveusorhateus}
k_2&=&\frac{8 C^5}{5} (1-2 C)^2 [2+2C(y_R-1)-y_R] \times \nonumber \\
&& \{ 2C[6-3y_R+3C(5y_R-8)]+\nonumber \\
&&4C^3[13-11y_R+C(3y_R-2)+ 2C^2(1+y_R)] + \nonumber \\
&&3 (1-2C)^2[2-y_R+2C(y_R-1)] \ln(1-2C) \}^{-1}  , \ \ \ \ \ \
\end{eqnarray}
with the compactness $C=M/R$. The quantity
$y_R\equiv y(R)$ on the other hand is obtained by solving the differential equation for $y(r)$ coming from the line element of the linearized metric, described in greater detail in \cite{1967ApJ...149..591T,Hinderer:2007mb,Damour:2009vw}
\begin{eqnarray}\nonumber
 ry'(r)&+&y(r)^2+r^2 Q(r) \\ \label{dgl_del_czz}
 &+&y(r)e^{\lambda(r)}\left[1+4\pi r^2(p(r)-\epsilon(r) )\right] = 0
 ,
 \label{y}
\end{eqnarray}
with 
\begin{eqnarray}\label{leunen}
Q(r)=4\pi e^{\lambda(r)}\left(5\epsilon(r)+9p(r)+\frac{\epsilon(r)+p(r)}{c_s^2(r)}\right)\\ \nonumber
-6\frac{e^{\lambda(r)}}{r^2}-(\nu'(r))^2, 
\end{eqnarray}
the metric functions from general relativity
\begin{eqnarray}
e^{\lambda(r)}&=&\left(1-\frac{2m(r)}{r} \right)^{-1},\\ 
\nu'(r)&=&2e^{\lambda(r)}\frac{m(r)+4\pi r^3 p(r)}{r^2}
\end{eqnarray}
and $c_s(r)^2=d p/ d \epsilon$ as the speed of sound squared \cite{Hinderer:2007mb,Hinderer:2009ca,Postnikov:2010yn}. 
The boundary condition of eq.~(\ref{y}) is $y(0)$=2, which implies no deformation at all in the center of the star \cite{zacchi_tdf}.\\ 
At the surface of a selfbound star $c_s^2 \rightarrow 0$ and the denominator in eq.~(\ref{leunen}) blows up. 
The density discontinuity leads to an extra expression just below the surface of the star \cite{Damour:2009vw,Hinderer:2009ca,Postnikov:2010yn}. 
One has to substract  
\begin{equation}\label{correctionsunit}
 y_s=\frac{4 \pi r^ 3 \epsilon_s(r)}{m(r)}
\end{equation} 
from eq.~(\ref{y}) with $\epsilon_s(r)$ being the value of the energy density just below the surface. 
The dimensionless tidal deformability $\Lambda$ is 
\begin{equation}\label{loveandloveismierda}
\Lambda=\frac{2k_2}{3 C^5},
\end{equation}
usually solved simultaneously with
the TOV equations \cite{Hinderer:2007mb,Hinderer:2009ca,Postnikov:2010yn}. The actual value of a 1.4$M_{\odot}$ star has to be in a range  $\Lambda=300^{+420}_{-230}$ \cite{TheLIGOScientific:2017qsa,Abbott:2018exr}. 

\section{Results}\label{resultini}
In this section we present our results for varying sigma meson mass $m_{\sigma}$, repulsive vector coupling $g_{\omega}$ and different values of the Bag constant $B^{1/4}$. At the end of this section we compare the results from the extended mean field approximation (eMFA) with the mean field approximation (MFA). The constituent quark mass is held fixed at $m_q=300$~MeV, which is roughly one third of the nucleon mass. The standard parameter set is $m_q=300$~MeV, $m_{\sigma}=600$~MeV, $g_{\omega}=3.5$ and  $B^{1/4}=80$~MeV, in the following denoted as reference set.
\subsection{Variation of the sigma meson mass}
\begin{figure}[H]
\center
\includegraphics[width=1.0\columnwidth]{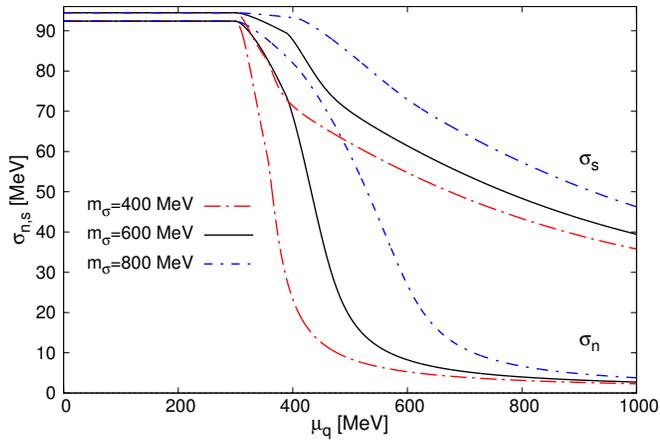}
\caption{\textit{The solutions of the condensate equations $\sigma_n$ and $\sigma_s$ in the extended mean field approximation as a function of the quark chemical potential $\mu_q$ for different values of $m_{\sigma}$. The other parameters are $m_q=300$~MeV, $g_{\omega}=3.5$ and  $B^{1/4}=80$~MeV.}}
\label{fig:msigmavarsigmafields}
\end{figure}

\begin{figure}[H]
\center
\includegraphics[width=1.0\columnwidth]{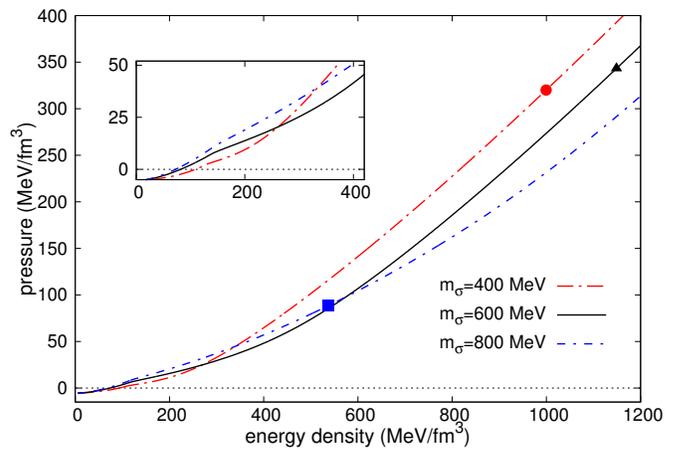}
\caption{\textit{The EoSs for different values of the sigma meson mass $m_{\sigma}$. The inlaid figure accentuates the behaviour at rather low energies, resulting in nontrivial mass-radius sequences. The symbols indicate where the maximum mass star is located on the respective EoS. The other parameters are $m_q=300$~MeV, $g_{\omega}=3.5$ and  $B^{1/4}=80$~MeV.}}
\label{fig:msigmavareos}
\end{figure}

\begin{figure}[H]
\center
\includegraphics[width=1.0\columnwidth]{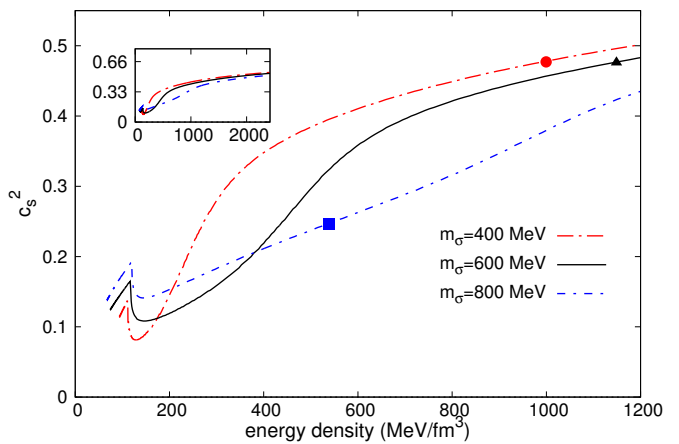}
\caption{\textit{The speed of sound for different values of the sigma meson mass $m_{\sigma}$. The symbols indicate the maximum mass star. The inlaid figure shows that the speed of sound approaches $c_s^2=0.5$ for large energy densities. The other parameters are $m_q=300$~MeV, $g_{\omega}=3.5$ and  $B^{1/4}=80$~MeV.}}
\label{fig:msigmavarsos}
\end{figure}

\begin{figure}[H]
\center
\includegraphics[width=1.0\columnwidth]{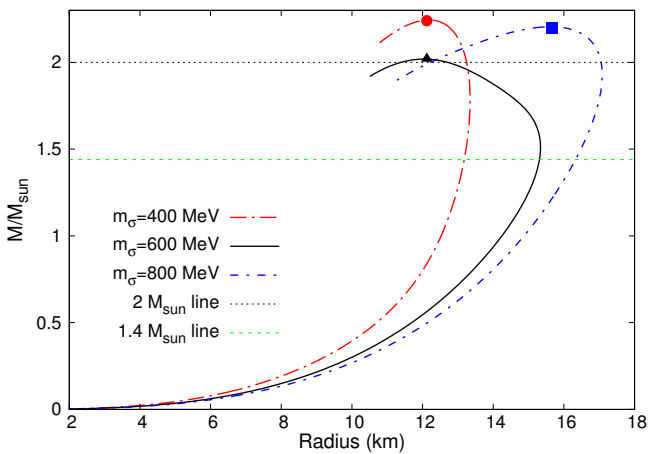}
\caption{\textit{The mass radius relation for different values of the sigma meson mass $m_{\sigma}$. For $m_{\sigma}=400$~MeV the maximum mass is 2.24$M_{\odot}$ at 12.12~km, for $m_{\sigma}=600$~MeV 2$M_{\odot}$ are reached at the same radius, whereas for $m_{\sigma}=800$~MeV  the maximum mass is 2.2$M_{\odot}$ at 15.65~km radius. These values are also listed in Tab.~\ref{wirseallejarjarbings}. The other parameters are $m_q=300$~MeV, $g_{\omega}=3.5$ and  $B^{1/4}=80$~MeV.}}
\label{fig:msigmavarmrr}
\end{figure}

\begin{figure}[H]
\center
\includegraphics[width=1.0\columnwidth]{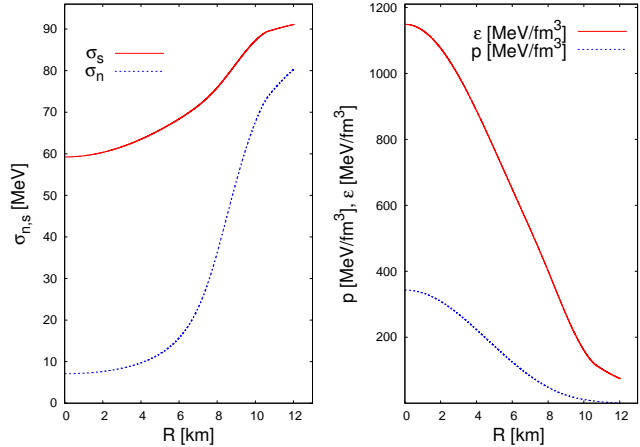}
\caption{\textit{The radial profile of the maximum mass star with with $M=2.02M_{\odot}$ at R=12.12~km (the black triangle) from the standard parameter set $m_q=300$~MeV, $m_{\sigma}=600$~MeV, $g_{\omega}=3.5$ and  $B^{1/4}=80$~MeV. The left figure displays the nonstrange- and the strange $\sigma$ condensate as a function of the stars radius R. The right figure shows the pressure $p$ and energy density $\epsilon$ as a function of the stars radius R.}}
\label{fig:radial_profile} 
\end{figure}

Figure \ref{fig:msigmavarsigmafields} shows the solutions of the scalar condensate equations, eqs.~(\ref{condensates}), as a function of quark chemical potential $\mu_q$ for the scalar fields $\sigma_n$ and $\sigma_s$. For all our choices of $m_{\sigma}$ a crossover transition is found. For larger values of $m_\sigma$ the restoration of chiral symmetry happens at larger quark chemical potential $\mu_q$ than for a lower mass of the sigma meson. 
A first order phase transition in the eMFA within this parameter choice is not found. 
Chiral symmetry is not entirely restored, because of the explicit breaking of chiral symmetry \cite{Lenaghan:2000ey,Zacchi:2015lwa}.
Figure \ref{fig:msigmavareos} shows the EoSs for three different values of the sigma meson mass, 400~MeV, 600~MeV and 800~MeV. 
The symbol on a particular EoS denotes the corresponding location of the maximum mass star on the EoS.
The red circle marks the maximum mass star on the EoS for $m_{\sigma}=400$~MeV. The black triangle is the maximum mass star on the reference set and the blue square is the representative maximum mass star for $m_{\sigma}=800$~MeV. 
The corresponding vacuum parameters are listed in Tab. \ref{wirsindallebluna}. 
Due to the implementation of the fermion vacuum term the behaviour of the grand potential,  eq.~(\ref{dumblidorr}), i.e. the resulting EoS is highly nonlinear. The appropriate EoSs stiffen with increasing sigma meson mass $m_{\sigma}$ for $p\leq 100~\rm{MeV/fm^3}$, which is in contrast to the MFA case  \cite{Zacchi:2015lwa}. 
For $p\geq 100~\rm{MeV/fm^3}$ the EoSs stiffen with decreasing sigma meson mass $m_{\sigma}$ as in the MFA .
The inlaid figure in Fig.~\ref{fig:msigmavareos} shows the crossing of the EoSs where the EoS for $m_{\sigma}=400~$MeV crosses the $m_{\sigma}=600~$MeV EoS at $p=20~$MeV/fm$^3$ and then the $m_{\sigma}=800~$MeV EoS at $p=35~$MeV/fm$^3$.
The corresponding speed of sound $c_s^2$ for the different choices of $m_{\sigma}$ can be seen in Fig.~\ref{fig:msigmavarsos}. The inlaid figure displays that the speed of sound approaches $c_s^2=0.5$ at large energy densities.  
The EoS for $m_\sigma=400$~MeV generates the highest values 
of the speed of sound $c_s^2$ at a given energy density for values of the energy density $\epsilon \geq 220$~$\rm{MeV/fm^3}$. This feature results from the stiffness of the EoS, see also Fig.~\ref{fig:msigmavareos}. 
$c_s^2$ influences the solution of the differential equation, eq.~(\ref{dgl_del_czz}), via the quantity $Q(r)$, eq.~(\ref{leunen}).  Thereby $k_2$ and eventually the tidal deformability $\Lambda$, eq.~(\ref{loveandloveismierda}), are influenced by the EoS. 
The mass radius relations can be seen in fig.~\ref{fig:msigmavarmrr}, where for our parameter choices   $2M_{\odot}$ are possible \cite{Demorest:2010bx,Antoniadis:2013pzd,Fonseca:2016tux,Cromartie:2019kug}, indicated by the upper horizontal line. The lower horizontal line indicates $1.4M_{\odot}$. 
For the lowest value of $m_{\sigma}=400$~MeV chosen in this article, the mass radius relation is more compact than for the other choices of $m_{\sigma}$. This feature is different in the MFA \cite{Zacchi:2015lwa} and  results from the nontrivial behaviour of the EoSs discussed for Fig.~\ref{fig:msigmavareos}. At $m_{\sigma}=400$~MeV the radius of a $1.4M_{\odot}$ star is $R_{1.4M_{\odot}}=13.14$~km, whereas  for $m_{\sigma}=600$~MeV $R_{1.4M_{\odot}}=15.27$~km and for $m_{\sigma}=800$~MeV $R_{1.4M_{\odot}}=16.22$~km, see also Tab.~\ref{wirseallejarjarbings}. With increasing $m_{\sigma}$ the configurations become less compact, but nontheless has the mass radius relation for $m_{\sigma}=600$~MeV the smallest maximum mass of $2.02M_{\odot}$.   
This value is at roughly the same radius $R=12.12$~km as the mass radius relation for $m_{\sigma}=400$~MeV with a maximum value of $2.24M_{\odot}$. The maximum mass for $m_{\sigma}=800$~MeV is $2.2M_{\odot}$ at a radius $R=15.65$~km. It is however not seen in the MFA that the maximum masses for two different values of $m_{\sigma}=400$~MeV and $m_{\sigma}=800$~MeV are nearly equal at different radii. 
This peculiarity can be explained with the nontrivial behaviour of the EoSs for $p \leq 100$~$\rm{MeV/fm^3}$, see inlaid figure in Fig.~\ref{fig:msigmavareos}.
Fig.~\ref{fig:radial_profile} shows the radial profile of the maximum mass star with $M=2.02M_{\odot}$ at R=12.12~km from the standard parameter set, i.e. the black triangle in figs.~\ref{fig:msigmavareos}, \ref{fig:msigmavarsos} and  \ref{fig:msigmavarmrr}. The left figure displays the nonstrange $\sigma_n$ and the strange $\sigma_s$ condensate as a function of the stars radius R. The right figure shows the pressure $p$ and the energy density $\epsilon$ as a function of the stars radius R. The curves are rather smooth because the phase transition is a crossover. 
The nonstrange $\sigma_n$ condensate has a value below 10~MeV/fm$^3$ in the center of the star at $R=0$, which is a magnitude smaller than the value of the chirally broken phase $f_{\pi}=92.4$~MeV.
However, chiral symmetry is not fully restored in the center of the $2.02M_{\odot}$ star as the strange $\sigma_s$ condensate at $R=0$ is only  slightly below 60~MeV for a vacuum expectation value of $94.47$~MeV. 


\subsection{Comparison with the mean field approximation}

The stars on the mass radius relation for a larger value of the repulsive coupling parameter $g_{\omega}$ have larger maximum masses and also larger radii,which is qualitatively known from the MFA case \cite{Weissenborn:2011ut,Zacchi:2015lwa}, see also the corresponding values in 
Tab.~\ref{wirseallejarjarbings} for the eMFA and Tab.~\ref{wirseallejarjarbingsmachtmeindingswiediesphinx} for the MFA.\\ 
The vacuum pressure constant $B^{1/4}$ drops out in the equations of motion, eqs.~(\ref{condensates}) and (\ref{tree_level_baby}) respectively. 
Smaller values of $B^{1/4}$ have essentially the same effect on the mass radius relation as a larger repulsive coupling $g_{\omega}$: Maximum masses and radii become larger.\\ 
\begin{figure}[H]
\center
\includegraphics[width=1.0\columnwidth]{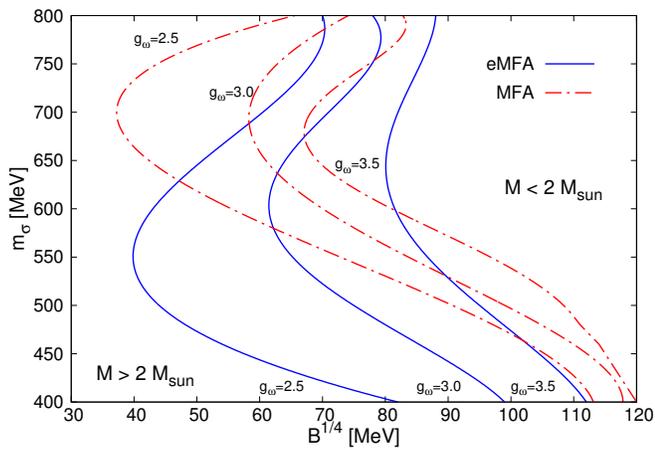}
\caption{\textit{Contour lines of maximum masses of 2$M_{\odot}$ in the $m_{\sigma}$ vs. $B^{1/4}$ plane for fixed $2.5 \leq g_{\omega} \leq 3.5$.  Smaller values of the repulsive coupling $g_{\omega}$ accompanied with a relatively small value of the vacuum pressure $B^{1/4}$ yields a maximum mass of $2M_{\odot}$ in the mass radius relation. This holds for the MFA (dashed lines) and for the eMFA (continuous lines).}}
\label{fig:kontour_1}
\end{figure}
Figure~\ref{fig:kontour_1} shows contour lines of $2M_{\odot}$ maximum mass for the MFA- and the eMFA case at fixed $g_{\omega}=2.5, 3$ and $3.5$ in the $m_{\sigma}$ vs. $B^{1/4}$ plane. 
For a parameter choice on the right hand side of a particular contour line, the maximum mass of the corresponding mass radius relation is smaller than $2M_{\odot}$, on the left hand side consequently  larger than $2M_{\odot}$. 
Furthermore, smaller values of the repulsive coupling $g_{\omega}$ together with a relatively small value of the vacuum pressure constant $B^{1/4}$ yield maximum masses of $\geq 2M_{\odot}$ in the mass radius relation \cite{Zacchi:2015lwa,Zacchi:2016tjw}. This statement holds for the MFA- and for the eMFA case.\\ 
The contour lines of the MFA and the eMFA seem somehow to be shifted vertically with a difference of $m_{\sigma}\simeq 150$~MeV for $g_{\omega}=2.5$, 
$m_{\sigma}\simeq 100$~MeV for $g_{\omega}=3.0$ and $m_{\sigma}\simeq 50$~MeV for $g_{\omega}=3.5$.
The difference in the shift in $m_{\sigma}$ becomes smaller for larger values of $g_{\omega}$.  
For a certain $m_{\sigma} \leq 620$~MeV, larger values of $B^{1/4}$ are allowed for $2M_{\odot}$ in the MFA compared to the eMFA, resulting in more compact mass radius relations in the MFA. Recall that smaller values of $B^{1/4}$ generate rather larger radii, so that denser stars yield smaller values of the tidal deformability parameter $\Lambda\propto C^{-5}$, see eq.~(\ref{loveandloveismierda}). For at least $2M_{\odot}$ in the eMFA with values $m_{\sigma} \geq 620$~MeV a rather large value of $B^{1/4}$ is necessary. The 
mass radius configurations however turn out to have already a too large radius for a small tidal deformability parameter $\Lambda \leq 720$, see also Tab.~\ref{wirseallejarjarbings} for the eMFA and Tab.~\ref{wirseallejarjarbingsmachtmeindingswiediesphinx} for the MFA.\\

\begin{figure}[H]
\center
\includegraphics[width=1.0\columnwidth]{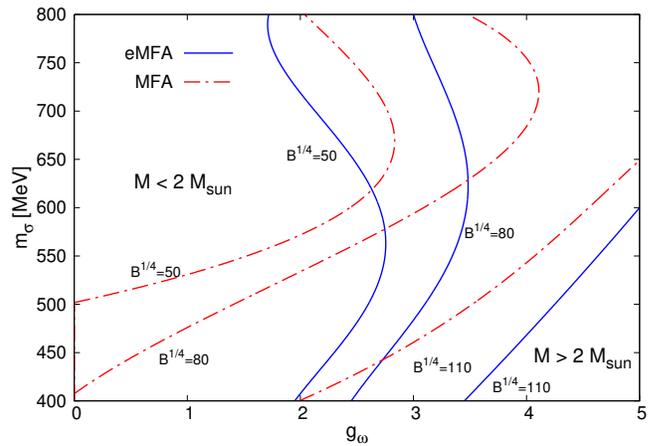}
\caption{\textit{Contour lines of maximum masses of 2$M_{\odot}$ in the $m_{\sigma}$ vs. $g_{\omega}$ plane at fixed $50 ~MeV \leq B^{1/4} \leq 110~MeV$. In the MFA case (dotted lines) no repulsive coupling is necessary to generate two solar masses for low $m_{\sigma}$. The configurations in MFA are more compact compared to the eMFA case (continuous lines).}}
\label{fig:kontour_2}
\end{figure}

Figure \ref{fig:kontour_2} shows contour lines of maximum masses of $2M_{\odot}$ for the MFA and the eMFA cases at fixed $B^{1/4}$ in the $m_{\sigma}$ vs. $g_{\omega}$ plane. For a particular parameter choice the maximum mass is larger than $2M_{\odot}$ on the right hand side of a particular contour line and on the left hand side consequently smaller. Recall that a larger value of the repulsive coupling is needed for $2M_{\odot}$ at fixed vacuum pressure constant $B^{1/4}$.
Interesting to note is, that 
in the MFA case no repulsive coupling is necessary to generate two solar masses for rather small  $m_{\sigma}$, making the configurations more compact compared to the eMFA case. As already mentioned, more compact configurations are in favour of a low value of the tidal deformability parameter $\Lambda$.
\begin{figure}[H]
\center
\includegraphics[width=1.0\columnwidth]{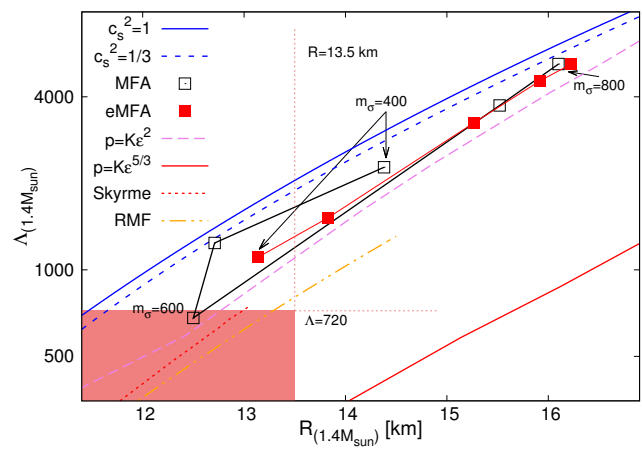}
\caption{\textit{The values of the tidal deformability parameter $\Lambda$ on a logarithmic scale for a $1.4M_{\odot}$ star as a function of the radius for a $1.4M_{\odot}$ star for various EoSs. The shaded area fulfills both constraints on either $R \leq 13.5$~km and $\Lambda \leq 720$.}}
\label{fig:LvsR}
\end{figure}
Figure \ref{fig:LvsR} shows the value of the tidal deformability parameter $\Lambda$ according to eq.~(\ref{loveandloveismierda}) on a logarithmic scale for a $1.4M_{\odot}$ star as a function of the radius of a $1.4M_{\odot}$ star. 
The shaded area fulfills both constraints on either $R \leq 13.5$~km 
\cite{Abbott:2018exr,Most:2018hfd,Annala:2017llu,De:2018uhw,Kumar:2017wqp,Fattoyev:2017jql,Malik:2018zcf,Hornick:2018kfi} 
and $\Lambda \leq 720$ \cite{Abbott:2018wiz}.
The $c_s^2$ curve corresponds to the MIT Bag model EoS $p=c_s^2 \epsilon-4B$, and is obtained for different values of the vacuum pressure constant B. Interesting to note is that the curves for the MIT Bag model for $c_s^2=1$ and $c_s^2=1/3$ are relatively close together. This feature indicates that a particular function $\Lambda_{1.4M_{\odot}}(R_{1.4M_{\odot}})$ is rather independent on the speed of sound $c_s^2$, so that $c_s^2$ in eq.~(\ref{leunen}) plays a subdominant part. The curve for $c_s^2=1$ can also be seen as an upper limit due to causality.\\ 
The MFA and the eMFA cases are obtained by varying $m_{\sigma}$ in the standard parameter set. In the eMFA case the $\Lambda_{1.4M_{\odot}}$ values and the radii decrease linearily on the logarithmic scale with
decreasing $m_{\sigma}$ (see also Tab.~\ref{wirseallejarjarbings}). In the MFA case a minimum value $\Lambda=680$ at 12.51~km radius for $m_{\sigma}=600$~MeV is found (see also Tab.~\ref{wirseallejarjarbingsmachtmeindingswiediesphinx}). Smaller and larger values of $m_{\sigma}=600$~MeV lead to larger values of either $\Lambda_{1.4M_{\odot}}$ and the corresponding radius.
This feature may be explained due to a shift in the dominance of attractive and repulsive field contributions to the stiffness of the EoS, and has already be suspected and discussed in \cite{Zacchi:2015lwa}. 
The only paramter set which respects  the $2M_{\odot}$ limit, the $R_{1.4M_{\odot}}\leq 13.5$~km bound and the $\Lambda_{1.4M_{\odot}}\leq 720$ constraint, is the reference set in the MFA, and is hence located  within the shaded area in fig.~\ref{fig:LvsR}. \\
The inclusion of the fermion vacuum term in the SU(3) chiral quark meson model seems not to be compatible with astrophysical measurements and constraints.\\ 
To sort our results for the SU(3) quark meson model within other approaches, a Skyrme parameter approach taken from Zhou et al. \cite{Zhou:2019omw} and an relativistic mean field (RMF) model studied by Nandi et al. \cite{Nandi:2018ami} are included in fig.~\ref{fig:LvsR}. The RMF values correspond to the fit function $\Lambda_{1.4M_{\odot}}=1.53\cdot 10^{-5}(R_{1.4M_{\odot}}/km)^{6.83}$ \cite{Nandi:2018ami}. The values of these two approaches are located at smaller $\Lambda_{1.4M_{\odot}}$ at a given radius and lie well within the shaded area. Furthermore, the results for free Fermi gas EoSs $p=K\epsilon^{1+1/n}$ for various constants K, with $n=1$ and $n=3/2$ are also shown.
The values of $\Lambda_{1.4M_{\odot}}(R_{1.4M_{\odot}})$ for the interaction dominated EoS with $n=1$ lie in between the results of the quark matter EoSs and the Skyrme-~and RMF approach.
The results of the nonrelativistic EoS for $n=3/2$ may be seen as an lower limit for the function 
$\Lambda_{1.4M_{\odot}}(R_{1.4M_{\odot}})$ in Fig.~\ref{fig:LvsR}.
The values for $\Lambda_{1.4M_{\odot}}$ at a given radius of the 
hadronic EoSs are located above the values of the nonrelativistic EoS.\\ 
In general it seems, that stars composed of quark matter have larger values of the 
tidal deformability parameter $\Lambda$ for a 1.4$M_{\odot}$ star at a given radius 
than stars generated by hadronic EoSs. 
     \section{Conclusions}
We have studied the SU(3) quark meson model including the fermion vacuum term and have determined the vacuum parameters for different values of the sigma meson mass $m_{\sigma}$. 
The whole potential is independent on any renormalization scale  \cite{Lenaghan:1999si,Gupta:2011ez,Chatterjee:2011jd,Tiwari:2013pg}.\\
For all scalar meson masses in the eMFA a crossover transition in the condensates is found. 
In general it seems that in the MFA a first order phase transition is possible in a larger parameter space. 
For larger values of $m_\sigma$ the restoration of chiral symmetry happens at larger quark chemical potential $\mu_q$ \cite{Han:2018mtj},
which is also observed for larger values of $g_{\omega}$. The Bag constant $B^{1/4}$ on the other hand does not affect the condensates at all, since it drops out in the equations of motion.\\
The resulting equations of state (EoSs) have been used to calculate mass radius relations of compact stars. These mass radius relations have to respect the $2M_\odot$ limit
\cite{Demorest:2010bx,Antoniadis:2013pzd,Fonseca:2016tux,Cromartie:2019kug} 
and the constraints coming 
from the analysis of the tidal deformabilities of the
GW170817 neutron star merger event
\cite{Abbott:2018wiz,Abbott:2018exr,Most:2018hfd,Annala:2017llu,De:2018uhw,Kumar:2017wqp,Fattoyev:2017jql,Malik:2018zcf}.
We compare our results from the extended mean field approximation (eMFA) with the mean field approximation (MFA). Our finding is that the $2M_{\odot}$ constraint can be fulfilled in both approaches in a rather wide parameter space, see e.g. \cite{Zacchi:2015lwa}, and that in the eMFA the mass radius relations are generally less compact resulting in larger values of the tidal deformability parameter $\Lambda$. This feature however implies that in the eMFA the constraints from GW170817 are not fulfilled, i.e. $\Lambda_{1.4M_{\odot}} \geq 720$. In the MFA a small parameter space is found where all considered restrictions are satisfied.\\ 
Within our parameter choice in the eMFA a smaller value of the sigma meson mass $m_{\sigma}$ is favoured for a rather compact mass radius relation. This is in contrast to the MFA, see e.g. \cite{Zacchi:2015lwa}, and may be explained via the additional term in the potential resulting from the fermion vacuum contribution. \\
A large repulsive coupling constant $g_{\omega}$ stiffens the EoS and hence enables the star to generate more pressure against gravity. The maximum mass and also the radius become consequently larger. These features may then result in larger values for the tidal deformability parameter $\Lambda \propto C^{-5}$, $C$ being the compactness. Incidentally, 
this holds vice versa, i.e. small respulsive couplings $g_{\omega}$ imply rather small tidal deformabilities, but the $2M_{\odot}$ constraint may not be fulfilled.\\ 
The EoSs substantially soften when increasing the vacuum pressure constant $B^{1/4}$ so that for values $B^{1/4}\geq 110$~MeV maximum masses of $2M_{\odot}$ are difficult to obtain. Smaller values of $B^{1/4}$ on the other hand have essentially the same effect on the mass radius relation as a larger repulsive coupling $g_{\omega}$, i.e. maximum masses and radii become larger and consequently also the tidal deformability parameter $\Lambda$.\\
To sort our results we compare our findings for the tidal deformability parameter at 1.4$M_{\odot}$, $\Lambda_{1.4M_{\odot}}(R_{1.4M_{\odot}})$, with the results for a constant speed of sound (linear) EoS with $c_s^2=1$, where $M\propto R^3$ and which can be seen as an upper limit due to causality. As a lower limit we introduce a nonrelativistic polytropic EoS with the polytropic index $\Gamma=5/3$ where $M \propto R^{-3}$. The case $\Gamma=2$ corresponds to an interaction dominated EoS where $R\simeq$~const., independent on the mass of the stars on the mass radius relation.  
In between these results we find the results of the MFA and eMFA cases relatively close to the constant speed of sound EoSs and the polytrope $n=2$.
For further comparison we also 
took a Skyrme parameter approach taken from Zhou et al. \cite{Zhou:2019omw} and an relativistic mean field (RMF) model studied by Nandi et al. \cite{Nandi:2018ami}. Their results are located slightly below the values of the polytropic $\Gamma=2$ case and centrally in between the results from the constant speed of sound- and the nonrelativistic polytropic EoS.



In general it seems, that quark matter stars have larger values of the tidal deformability parameter $\Lambda$ at 1.4$M_{\odot}$ than hadronic stars. 
Taking into account more interactions among quarks could lower the value of the tidal deformability parameter $\Lambda_{1.4M_{\odot}}(R_{1.4M_{\odot}})$, because interactions among quarks may decrease the value of $\Lambda_{1.4M_{\odot}}$ at a given radius, see Fig.\ref{fig:LvsR}.\\
A first order chiral phase transition yielding twin stars as in the MFA, see e.g. \cite{Zacchi:2016tjw}, was not found in the eMFA.  Compared to the mean field case, the EoSs show a smooth behaviour due to the additional term in the potential coming from the fermion vacuum term. 
The MFA and the eMFA, however, yield stars with $\geq 2M_{\odot}$ and radii $\leq 13.5$~km at $1.4M_{\odot}$, but only in the MFA the additional $\Lambda \leq 720$ limit was fulfilled for one set of parameters, i.e. for $m_{\sigma}=600$~MeV. This is due to the fact that the MFA approach yields more compact mass radius relations than the eMFA case. In the eMFA the $2M_{\odot}$ limit is fulfilled for the compactest mass radius configuration at $m_{\sigma}=400$~MeV with a radius of 13.14~km, but the value of $\Lambda_{1.4M_{\odot}}(R_{1.4M_{\odot}})=1107$ is not compatible with GW170817.


\begin{table*}
\begin{center}
\begin{tabular}{|c|c|c||c|c||c|c|c|c|}
\hline\hline 
 $m_{\sigma}$ [MeV] & $g_{\omega}$ & $B^{1/4}$ [MeV]  & $R_{1.4_{M\odot}}$ [km]& $\Lambda_{1.4_{M\odot}}$ & $R_{max}$ [km] & $M_{max}$ $[M_{\odot}]$ & $p_{max} [MeV/fm^3]$ & $\epsilon_{max}$ $[MeV/fm^3]$\\
\hline
 400 & 3.5 & 80 & 13.14 &1107& 12.12 & 2.24 & 320.02 & 999.55\\
\cline{1-9}
 600 & 3.5 & 80 & 15.27 &3253& 12.12 & 2.02 & 343.21 & 1148.82 \\
\cline{1-9}
 800 & 3.5 & 80 & 16.22 &5184& 15.65 & 2.20   & 88.62 & 537.67\\ 
\cline{1-9}
 600 & 1 & 80 & 9.32 &47& 8.87 & 1.44  & 409.65 & 1829.86\\ 
\cline{1-9}
 600 & 6 & 80 & 18.06 &1079& 17.15 & 2.82   & 128.35 & 509.77 \\ 
\cline{1-9}
 600 & 3.5 & 50 & 22.57 &5250& 22.80 & 2.61 & 23.35 & 221.13 \\ 
 \cline{1-9}
 600 & 3.5 & 110 & 10.10 &66& 9.36 & 1.79 & 578.83 & 1658.17 \\ 

\hline \hline \hline
\end{tabular}
\caption{\textit{eMFA: The values of the radius at $1.4M_{\odot}$, the tidal deformability parameter $\Lambda_{1.4M_{\odot}}$, the maximum radius, maximum mass and the central pressure and central energy density for the maximum mass stars for the parameters $m_{\sigma}$, $g_{\omega}$ and $B^{1/4}$.}}
\label{wirseallejarjarbings}
\end{center}
\end{table*}
\begin{table*}
\begin{center}
\begin{tabular}{|c|c|c||c|c||c|c|c|c|}
\hline\hline 
 $m_{\sigma}$ [MeV] & $g_{\omega}$ & $B^{1/4}$ [MeV]  & $R_{1.4_{M\odot}}$ [km]& $\Lambda_{1.4_{M\odot}}$ & $R_{max}$ [km] & $M_{max}$ $[M_{\odot}]$ & $p_{max} [MeV/fm^3]$ & $\epsilon_{max}$ $[MeV/fm^3]$\\
\hline
 400 & 3.5 & 80 & 14.38 &2275&15.28  &2.70  &142.70  &568.15 \\
\cline{1-9}
 600 & 3.5 & 80 & 12.51  &680& 11.47 &2.03 & 330.56 &1143.22  \\
\cline{1-9}
 800 & 3.5 & 80 & 16.10  &4858& 15.88 &  2.09  &64.78  &480.79 \\ 
\cline{1-9}
 600 & 1 & 80 &10.74  &389& 10.45 & 1.71  &206.77  &1091.38 \\ 
\cline{1-9}
 600 & 6 & 80 & 18.00 &9815& 14.86 &2.58   &251.90  &761.90  \\ 
\cline{1-9}
 600 & 3.5 & 50&21.92 &25300  &14.5& 2.19 &228.14  &878.17    \\ 
 \cline{1-9}
 600 & 3.5 & 110 & 10.08 &243&9.61  & 1.82 &467.07  &1493.87  \\ 

\hline \hline \hline
\end{tabular}
\caption{\textit{MFA: The values of the radius at $1.4M_{\odot}$, the tidal deformability parameter $\Lambda_{1.4M_{\odot}}$, the maximum radius, maximum mass and the central pressure and central energy density for the maximum mass stars for the parameters $m_{\sigma}$, $g_{\omega}$ and $B^{1/4}$.}}
\label{wirseallejarjarbingsmachtmeindingswiediesphinx}
\end{center}
\end{table*}
\begin{acknowledgments}
The authors thank Jan-Erik Christian for fruitful discussions on the tidal deformability. 
J.S. and A.Z. acknowledge support from the Hessian LOEWE initiative 
through the Helmholtz International Center for FAIR (HIC for FAIR).
\end{acknowledgments}
\bibliography{neue_bib}
\bibliographystyle{apsrev4-1}
\end{document}